\documentclass[twocolumn,showpacs,preprintnumbers,amsmath,amssymb]{revtex4}

\usepackage{graphicx}
\usepackage{dcolumn}
\usepackage{bm}

\begin{document}

\title{High sensitivity optical Faraday-magnetometry with intracavity electromagnetically induced transparency}

\author{Qiaolin Zhang$^{1}$}
\author{Hui Sun$^{1}$}\thanks{hsun@siom.ac.cn}
\author{Shuangli Fan$^{1}$}
\author{Hong Guo$^{2}$}\thanks{hongguo@pku.edu.cn}

\affiliation{$^{1}$School of Physics and Information Technology,
Shaanxi Normal University, Xi'an 710062, China}
\affiliation{$^{2}$State Key Laboratory of Advanced Optical
Communication Systems and Networks, School of Electronics
Engineering and Computer Science, and Center for Quantum Information
Technology, Peking University, Beijing 100871, China}

\date{\today}

\begin{abstract}
We suggest a multiatom cavity quantum electrodynamics system for the
weak magnetic field detection based on Faraday rotation with
intracavity electromagnetically induced transparency. Our study
demonstrates that the collective coupling between the cavity modes
and the atomic ensemble can be used to improve the sensitivity. With
single probe photon input, the sensitivity is inversely proportional
to the number of atoms, and the sensitivity with
0.7(5)~nT/$\sqrt{\rm Hz}$ could be attained. With multiphoton
measurement, our numerical calculations show that the magnetic field
sensitivity can be improved to 4.7(9)~fT/$\sqrt{\rm Hz}$.
\end{abstract}

\maketitle

\section{introduction}

The ability to detect magnetic field by optical means with high
sensitivity~\cite{budker-np-2007,edelstein-jp-2007,taylor-np-2008}
is a key requirement for a wide range of practical applications
ranging from geology and medicine to mineral exploration and
defense. A particularly important application is magnetic resonance
imaging. A variety of techniques including a superconducting quantum
interference device (SQUID)~\cite{romalis-mt-2011}, cavity
optomechanical~\cite{forstner-prl-2012}, negatively charged
nitrogen-vacancy (NV) centers in
diamond~\cite{jensen-prl-2014,xia-pra-2015}, and Bose-Einstein
condensate (BEC)~\cite{eto-pra-2013,muessel-prl-2014} have been
suggested for extremely high sensitive optical magnetometry. There
are two ways for sensing of magnetic fields. One is based on light
absorption at a magnetic resonance, for example $^{4}$He atomic
magnetometry by optical
pumping~\cite{macgregor-rsi-1987,wu-rsi-2015}. Another way to detect
magnetic field is achieved by making use of the changes of the index
refraction such as high sensitivity optical
magnetometry~\cite{scully-prl-1992,fleischhauer-pra-1994,fleischhauer-pra-2000,petrosyan-pra-2004}
based on electromagnetically induced transparency
(EIT)~\cite{marangos-jmo-1998,fleischhauer-rmp-2005}.

EIT technique enables one to control the absorption and dispersion,
changes the index refractive with cancelled absorption. With the
presence of the static magnetic field, the left- and right-circular
components of the linear polarized probe field drive different
transitions, and thus accumulate different phase shifts. As a result
the polarization direction is rotated, which is the so-called
Faraday rotation~\cite{budker-rmp-2002}. Large optical Faraday
rotation has been observed~\cite{atature-np-2007}, and has been
suggested to detect quantum fluctuation~\cite{chen-sr-2014}, atomic
filter~\cite{tao-ol-2015,yin-oe-2014}. The sensitivity of Faraday
rotation to weak magnetic fields naturally suggests it as a
magnetometry technique~\cite{budker-rmp-2002,petrosyan-pra-2004}.

When an ensemble of two-level atoms is placed inside an optical
cavity, the atom-cavity interaction strength can be enhanced to be
$g\sqrt{N}$, where $g$ is the single-atom-cavity coupling strength
and $N$ is the number of atoms in the cavity mode. Due to the strong
optical confinement and small mode volumes, the optical cavity
provide an excellent platform for strong light-matter interactions
allowing for vacuum induced
transparency~\cite{tanji-suzuki-science-2011} and all-optical
transistor~\cite{chen-science-2013}. The resonator response is
consequently drastically modified, resulting in substantial
narrowing of spectral
features~\cite{lukin-ol-1998,wu-prl-2008,geabanacloche-pra-2008,peng-lpl-2014}.
The cavity-enhanced Faraday rotation in NV centers in diamond can
push the sensitivity of microwave magnetometer into aT/$\sqrt{\rm
Hz}$~\cite{xia-pra-2015}. Furthermore, owing to the established
technology for microcavity fabrication, cavities may be an
attractive choice for miniaturized systems.

In the present paper, we suggest a composite atom-cavity system for
high sensitivity optical Faraday magnetometry based on intracavity EIT. In
doing so, we apply a linear polarized probe field to couple into the
cavity, and analyze the transmissions and phase shifts of its left-
and right-circular polarized components based on intercavity
EIT~\cite{lukin-ol-1998,albert-np-2011,dantan-pra-2012,zou-pra-2013,sharma-pra-2015}. 
The main idea is to combine cavity-enhanced Faraday rotation and intracavity EIT. Our study show
that the cavity-enhanced Faraday rotation supports the detection of
weak magnetic field, and resulting in cavity-enhanced sensitivity.
For single probe photon measurement, the sensitivity is inversely
proportional to the number of atoms. With multiphoton measurement,
the limit of sensitivity can be improved to 4.7(9)~fT/$\sqrt{\rm
Hz}$ from $0.7(5)$nT/$\sqrt{\rm Hz}$ with single photon measurement. Comparing with the microwave Faraday magnetometry proposed by Xia \textit{et.al.} in
Ref.\cite{xia-pra-2015}, the physical mechanism behinds our Faraday magnetometry is intracavity EIT, not the resonance absorption. The transparency peak can therefore be controlled by the driving field and the adjustable number of atoms confined in the cavity. Secondly, our Faraday magnetometry works in the optical wave, instead of microwave.

\begin{figure}
\includegraphics[width=8cm]{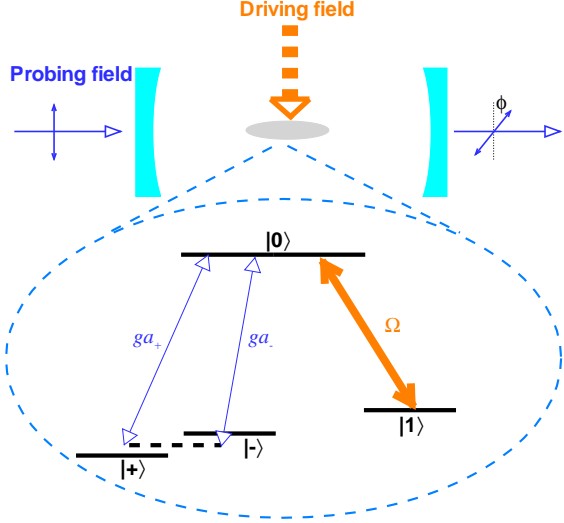}
\caption{(Color online) Schematics of setup and atom
configuration for weak magnetic field detection. The vertical linear
polarized probe field couples into the optical cavity, and then
transmits the cavity to the detector. The driving field $\Omega$ is
free propagating, and it is much larger than the size of the atomic
ensemble. We assume the cavity is symmetric such that the loss rates
of the cavity fields of the right and left mirrors are equal, i.e.,
$\kappa_{R}=\kappa_{L}=\kappa/2$. The decay rate of the excited state
$|0\rangle$ is denoted by $2\gamma$. We assume $g_{+}=g_{-}=g$ for
simplicity.}\label{fig:level}
\end{figure}

\section{Scheme and optical responses}

As depicted in Fig.~\ref{fig:level}, $N$ identical four-level atoms
in tripod configuration, which can be realized in $^{87}$Rb $D_{1}$ line
using hyperfine and Zeeman sublevel of the ground states, are
confined in the cavity. The three lower states $|+\rangle$,
$|-\rangle$, and $|1\rangle$ correspond to the levels
5$^{2}{\rm S}_{1/2}$ $|F=1,m_{F}=+1\rangle$, $|F=1,m_{F}=-1\rangle$,
and $|F=2,m_{F}=0\rangle$, respectively. We choose the state
5$^{2}{\rm P}_{1/2}$ $|F=1,m_{F}=0\rangle$ as the excited state
$|0\rangle$. The weak magnetic field lifts the degeneracy of the
Zeeman energy level 5$^{2}{\rm S}_{1/2}$ $|F=1,m_{F}=\pm1\rangle$,
and the energy shift is denoted by $\delta=g_{L}\mu_{B}B$ with
$g_{L}$ and $\mu_{B}=14.0$MHz$\cdot$mT$^{-1}$ being Lande $g$-factor
and Bohr magneton. A vertical (V) linear polarized probe field with
carrier frequency $\omega_{p}$ couples into the cavity, and its
left- and right-circular polarized components ($\sigma_{\pm}$) drive
the frequency-degenerated left- and right-circular polarized cavity
modes $(\{\hat{a}_{\pm}\})$. We denote the detuning between the
probe field with the cavity mode by $\Delta=\omega_{p}-\omega_{c}$
with $\omega_{c}$ being the cavity mode frequency. The transitions
$|\pm\rangle\leftrightarrow|0\rangle$ are driven by the cavity mode
$g_{\pm}\hat{a}_{\pm}$. A classical linear polarized driving field
with carrier frequency $\omega_{c}$ from free space couples the
transition $|1\rangle\leftrightarrow|0\rangle$. Thus the $\Lambda$
configuration, which is the heart of standard EIT, for cavity modes
$\hat{a}_{\pm}$ are formed.

In a rotating frame with the probe and driving field frequencies,
the interacting Hamiltonian for the coupled multiatom-cavity system has
the following form
\begin{eqnarray}
H=H_{a}+H_{af}+H_{f},\label{eq:hamiltonian-t}
\end{eqnarray}
in which
\begin{subequations}\begin{eqnarray}
&&\hspace{-0.8cm}H_{a}=\hbar\sum\limits_{i}(\Delta_{p}-\delta)\hat{\sigma}_{++}^{i}
+(\Delta_{p}+\delta)\hat{\sigma}_{--}^{i}
+\Delta_{d}\hat{\sigma}_{11}^{i},\label{eq:hamiltonian-a}\\
&&\hspace{-0.8cm}H_{af}=-\hbar\sum\limits_{i}\left(g_{+}\hat{a}_{+}\hat{\sigma}_{0+}^{i}
+g_{-}\hat{a}_{-}\hat{\sigma}_{0-}^{i}+\Omega\hat{\sigma}_{01}^{i}\right)+{\rm H.c.},\nonumber\\\label{eq:hamiltonian-af}\\
&&\hspace{-0.8cm}H_{f}=-\hbar\Delta\hat{a}_{+}^{\dagger}\hat{a}_{+}-\hbar\Delta\hat{a}_{-}^{\dagger}\hat{a}_{-},\label{eq:hamiltonian-f}
\end{eqnarray}\end{subequations}
where
$\Delta_{p}=\omega_{p}-[\omega_{0}-(\omega_{+}+\omega_{-})/2]$, and
$\Delta_{d}=\omega_{d}-(\omega_{0}-\omega_{1})$, are one photon
detunings. $\sigma_{\alpha\beta}^{i}$ ($\alpha,\beta=+,-,0,1$) is
the atomic operator for the $i$-th atom. $\hat{a}_{\pm}$
($\hat{a}_{\pm}^{\dagger}$) is the annihilation (creation) operator
of the cavity photons, and the cavity-atom coupling coefficient is
denoted by
$g_{\pm}=\mu_{\pm0}\sqrt{\omega_{c}/2\hbar\varepsilon_{0}V}$. As
usual, we denote the Rabi frequency of the classical driving field
by $2\Omega$. For simplicity, we assume $g_{+}=g_{-}=g$ and the
cavity is symmetric, i.e., $\kappa_{L}=\kappa_{R}=\kappa/2$.

In the following, we focus our attention on the transmission and the
Faraday rotation angle of the probe field. The responses of the two
circular polarized components of the probe field can be
characterized by intensity transmission coefficients $t_{\pm}$ and
the phase shifts $\phi_{\pm}$ of the corresponding components at the
output. Following the standard
processes\cite{qo-walls,zou-pra-2013}, we solve the Heisenberg
equations in steady-state and weak-field limit ($g\ll\Omega$).
The steady-state solutions read
\begin{eqnarray}
\frac{a_{\pm}^{\rm T}}{a_{\pm}^{\rm
in}}=t_{\pm}e^{i\phi_{\pm}}=\frac{\kappa}{\kappa-i\Delta-i\chi_{\pm}},\label{eq:tran}
\end{eqnarray}
where $\chi_{\pm}$ are the atomic susceptibilities, and they are
given by
\begin{eqnarray}
\chi_{\pm}=-i\frac{g^{2}N}{2(d_{j0}+|\Omega|^{2}/d_{j1})},\hspace{0.3cm}j=+,-,
\end{eqnarray}
in which $d_{\pm0}=i(\Delta_{p}\pm\delta)-\gamma$,
$d_{\pm1}=i(\Delta_{p}-\Delta_{d}\pm\delta)-\gamma'$. $2\gamma$ is
decay rate of the excited state $|0\rangle$, and $\gamma'$ is the
dephasing rate between the three lower states.

\begin{figure}
\includegraphics[width=8cm]{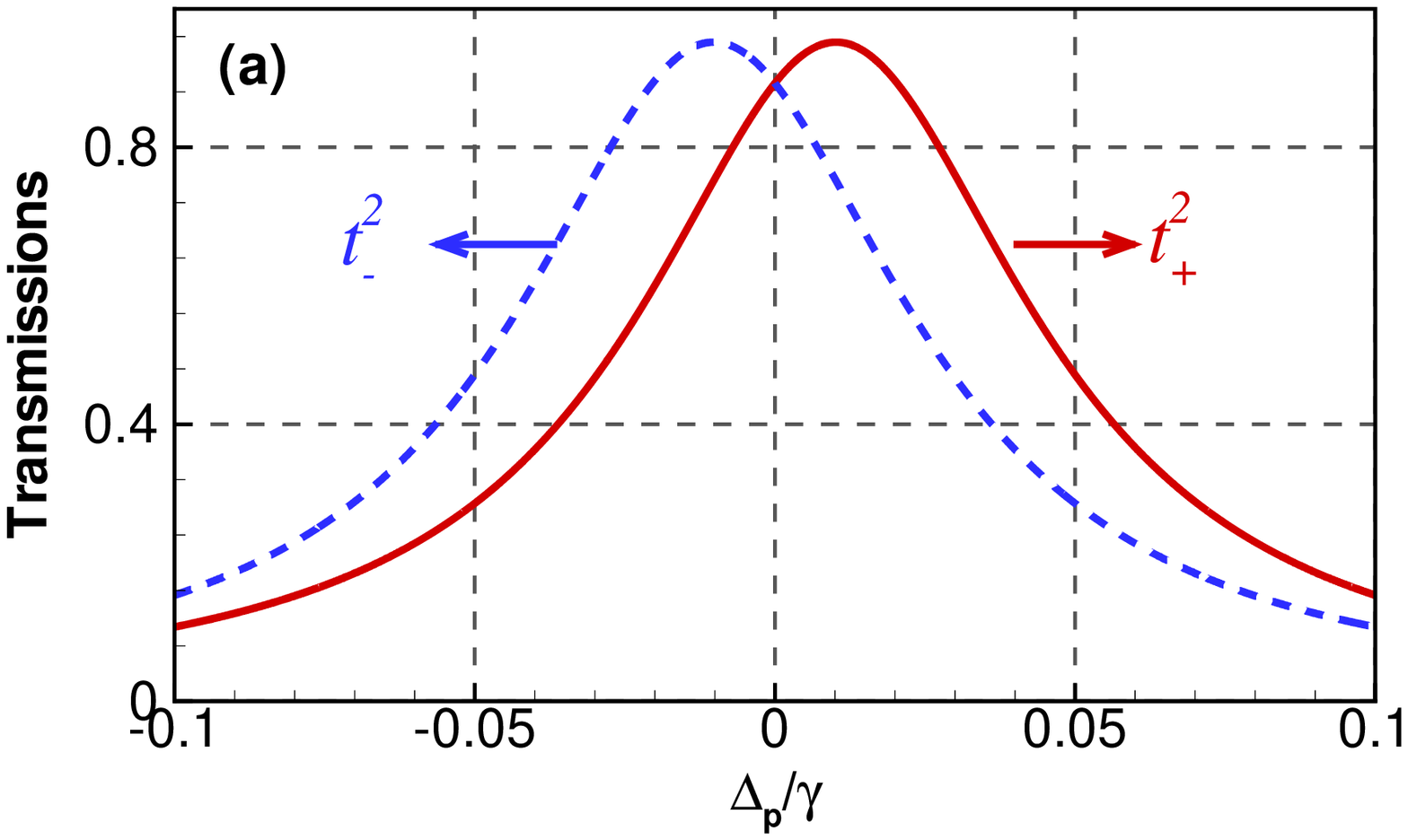}
\includegraphics[width=8cm]{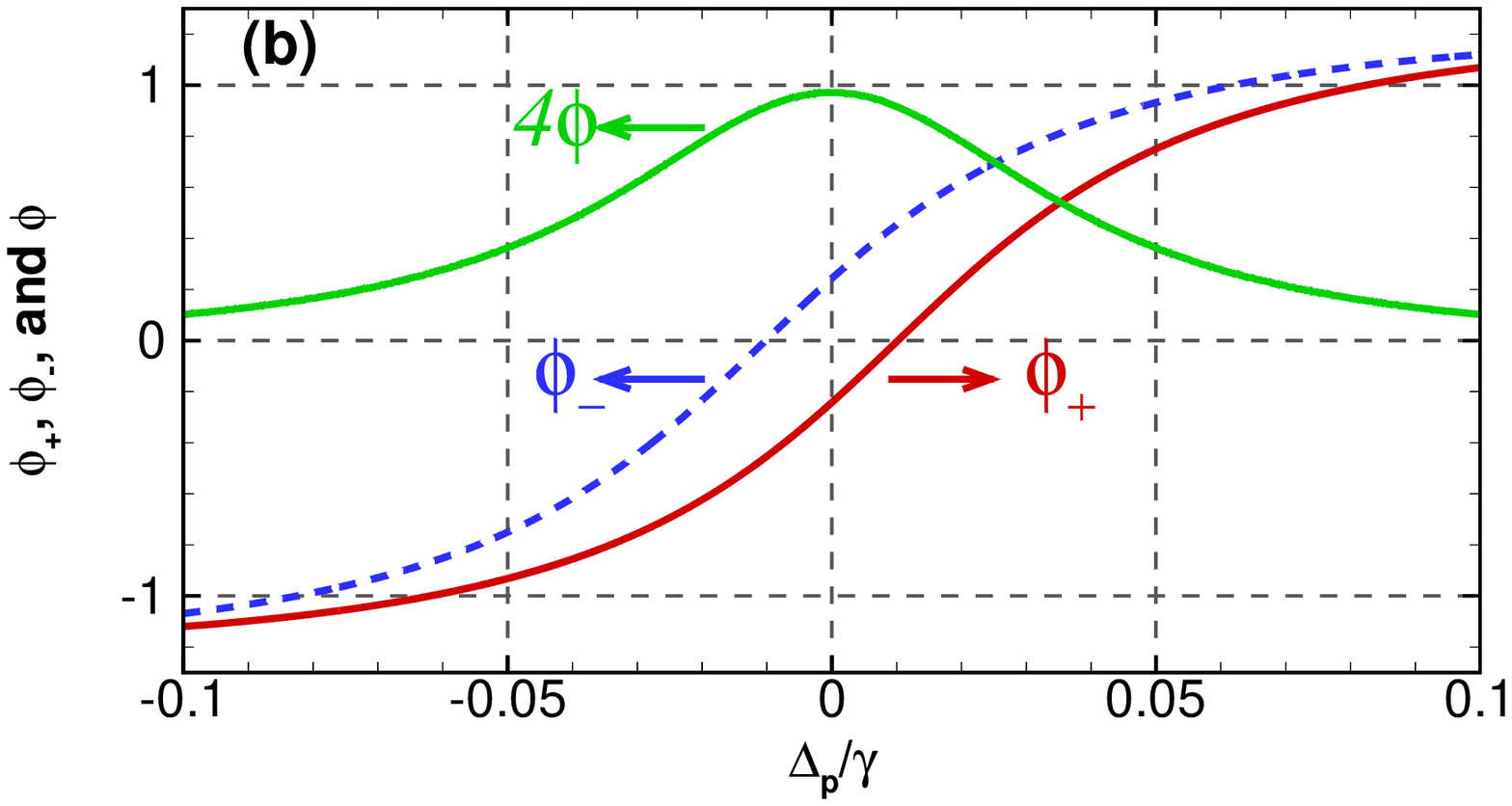}
\caption{(Color online) (a) The intensity ratio of the left and
right circular polarized cavity-transmitted probe field $t_{+}^{2}$
(solid curve) and $t_{-}^{2}$ (dashed curve) versus the probe
detuning $\Delta_{p}/\gamma$. (b) The phase shift of the
$\sigma_{+}$ and $\sigma_{-}$ cavity-transmitted probe field
$\phi_{+}$ (solid curve) and $\phi_{-}$ (dashed curve) and the
Faraday rotation angle $\phi$ (dotted curve) as functions of the
probe detuning $\Delta_{p}/\gamma$. The parameters are taken as
$g\sqrt{N}=10\gamma$, $\Omega=1\gamma$, $\Delta_{d}=0\gamma$,
$\kappa=2\gamma$, $\gamma'=10^{-3}\gamma$, $\Delta=0$, and
$\delta=10^{-2}\gamma$. The rotation angle $\phi$ is amplified by
four times for clarify.}\label{fig:tran-phase}
\end{figure}

In order to provide a detailed picture for weak magnetic field
detection, it is illustrative to present the intensity ratio of the
cavity transmitted probe fields $t_{\pm}^{2}$ and phase shifts
$\phi_{\pm}$ of the two circular components, and they are shown in
Figs.~\ref{fig:tran-phase}(a) and \ref{fig:tran-phase}(b) as
functions of the probe detuning $\Delta_{p}$. We take the relevant
parameters as $g\sqrt{N}=10\gamma$, $\Omega=1\gamma$,
$\Delta_{d}=\Delta=0\gamma$, $\kappa=2\gamma$,
$\gamma'=10^{-3}\gamma$, and $\delta=10^{-2}\gamma$. As presented in Fig.\ref{fig:tran-phase}(a), the probe transmissions of the two circular components of probe field are Lorentzian shaped around $\Delta_{p}=\pm\delta$, and the widths of the transparency peaks are substantially smaller than natural linewidth $\gamma$. This is different from the conventional
EIT~\cite{marangos-jmo-1998,fleischhauer-rmp-2005}. The weak
magnetic field lifts the degeneracy of Zeeman level $|\pm\rangle$,
and thus the positions of the transparency peaks are totally
determined by the strength of the magnetic field, i.e.,
$\Delta_{p}=\delta$ ($\Delta_{p}=-\delta$) for $\sigma_{+}$ ($\sigma_{-}$) component. With weak magnetic field presence, we have $t_{+}^{2}=t_{-}^{2}\approx0.89(6)$.
The vanishing absorptions ensure the applications of the composite
system for weak magnetic field detection at low-light level.

Within the narrow transparency peaks, the phase shifts of the two
components of the probe field $\phi_{\pm}$ are shown in
Fig.~\ref{fig:tran-phase}(b). Owing to intracavity EIT, the
dispersion curves are narrowed. As a result, the phase shifts
$\phi_{+}$ (solid curve) and $\phi_{-}$ (dashed curve) vary rapidly,
leading to the enhancement of the Faraday rotation angle
$\phi=(\phi_{-}-\phi_{+})/2$. This is the so called cavity-enhanced
Faraday rotation\cite{xia-pra-2015}. The Faraday rotation angle
$\phi$ versus probe detuning $\Delta_{p}$ is illustrated in
Fig.~\ref{fig:tran-phase}(b) with dotted curve. For the sake of
clarity, we amplify $\phi$ by four times. With this set of parameters, the effective half widths of transparency peaks and Faraday rotation angle are, respectively, $w_{t}\approx0.04\gamma$ and $\phi\approx0.26$~rad. Owing to the
cavity-enhanced Faraday rotation, the rotation angle $\phi$ is very
sensitive to the magnetic field, which makes the cavity-enhanced
sensitivity limit feasible.

In the regime of standard intracavity EIT ($\gamma\gamma'\ll\Omega^{2}$), and combining with the relation $g\sqrt{N}\gg\Omega$~\cite{fleischhauer-rmp-2005}, the effective half widths of the transparency peaks of the two circular components are~\cite{lukin-ol-1998,dantan-pra-2012}
\begin{eqnarray}
w_{t}\approx\gamma'+2\kappa\frac{\Omega^{2}}{g^{2}N}.\label{eq:width-transparent-peak}
\end{eqnarray}
The effective half widths are proportional to $(\Omega/g\sqrt{N})^{2}$, and they are much smaller than the bare cavity half width $2\kappa$ when $g\sqrt{N}\gg\Omega$. We assuming the atomic ensemble is cold, i.e., taking the dephasing rates of the three lower states as
$\gamma'=5\times10^{-4}\gamma$ such that the transparent condition
$\gamma\gamma'\ll\Omega^{2}$ can be satisfied~\cite{fleischhauer-rmp-2005}. Changing the parameters as $\Omega=0.5\gamma$ and
$g^{2}N=200\gamma^{2}$, one immediately obtains $w_{t}\approx5\times10^{-3}\gamma$. The Faraday rotation angle reach $\phi\approx0.18$~rad with $\delta=10^{-3}\gamma$. In the typical transparency peaks, the rapid changes in dispersion around two-photon resonance enhance the sensitivity of the Faraday rotation angle on the strength of the weak magnetic field.

In order to see the role of the cavity more clearly, we next calculate
the sensitivity to weak magnetic fields, which is the most important
characteristic of magnetometry. Applying the resonant conditions
($\Delta_{p}=\Delta_{d}=\Delta_{c}=0$) and neglecting the dephasing
rates ($\gamma'=0$), we have
\begin{eqnarray}
&&t_{\pm}=\frac{2\kappa(\delta'^{2}+\gamma^{2})^{1/2}}
{[(2\kappa\gamma-2\delta\delta'+g^{2}N)^{2}+4(\kappa\delta'+\delta\gamma)^{2}]^{1/2}},\\
&&\phi_{\pm}=\pm\frac{
g^{2}N\delta'-2\delta(\delta'^{2}+\gamma^{2})}{g^{2}N\gamma+2\kappa(\delta'^{2}+\gamma^{2})},
\end{eqnarray}
in which $\delta'=\delta-\Omega^{2}/\delta$. At the ``resonant point"
($\Delta_{p}=0$), the phase shifts of the two circular components of
the probe field are exactly equal and opposite in sign due to the
symmetry. In the limit of small magnetic field, the polarization
rotation angle can be reduced to
$\phi=(\phi_{-}-\phi_{+})/2=\phi_{-}\simeq(\delta/2\kappa)(g^{2}N/\Omega^{2})$,
which indicates clearly that the Faraday rotation can be enhanced by
the collective coupling of atoms with the cavity modes.

\section{Optical magnetometry and its sensitivity}

\begin{figure}
\includegraphics[width=8cm]{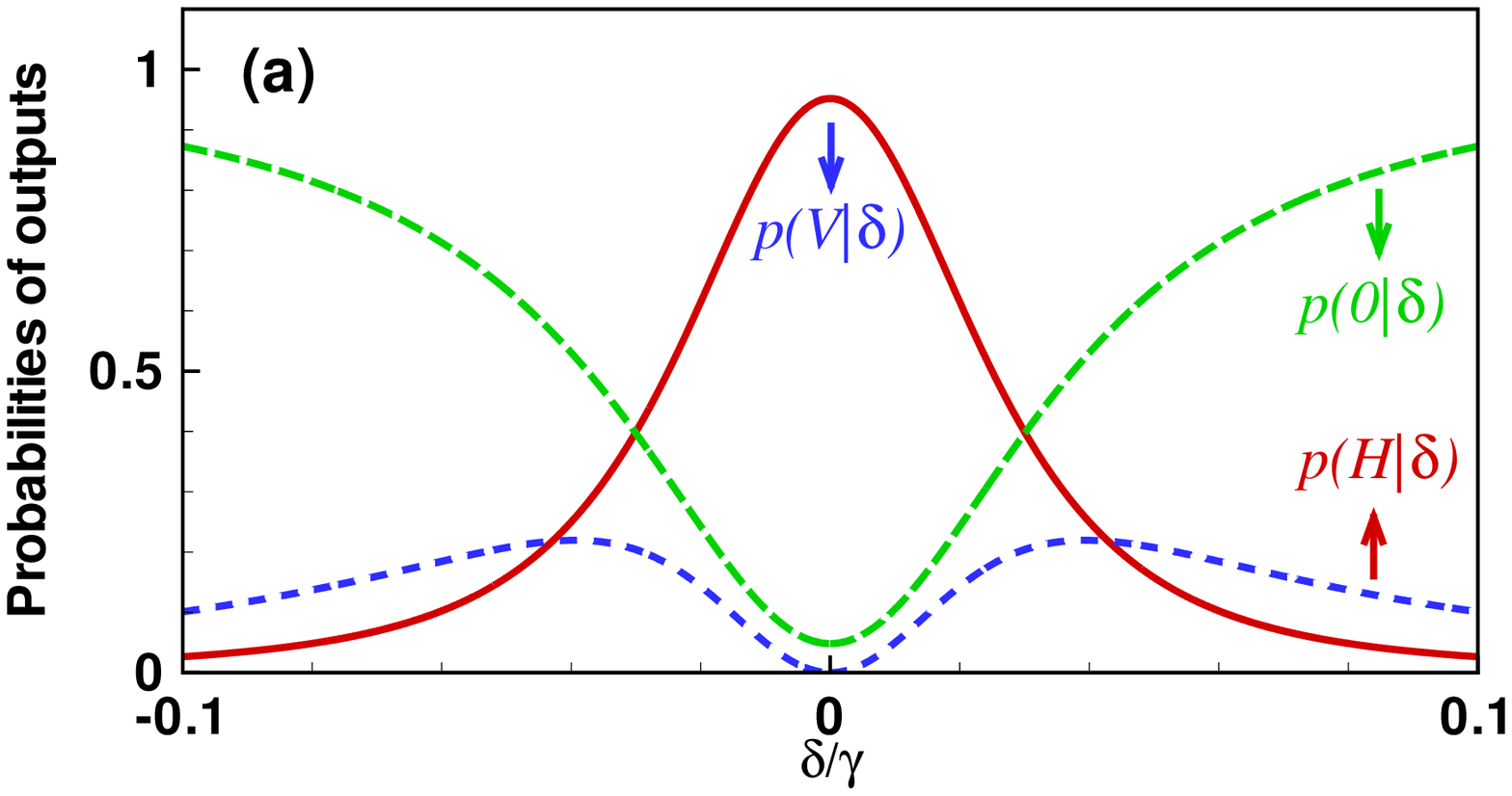}
\includegraphics[width=8cm]{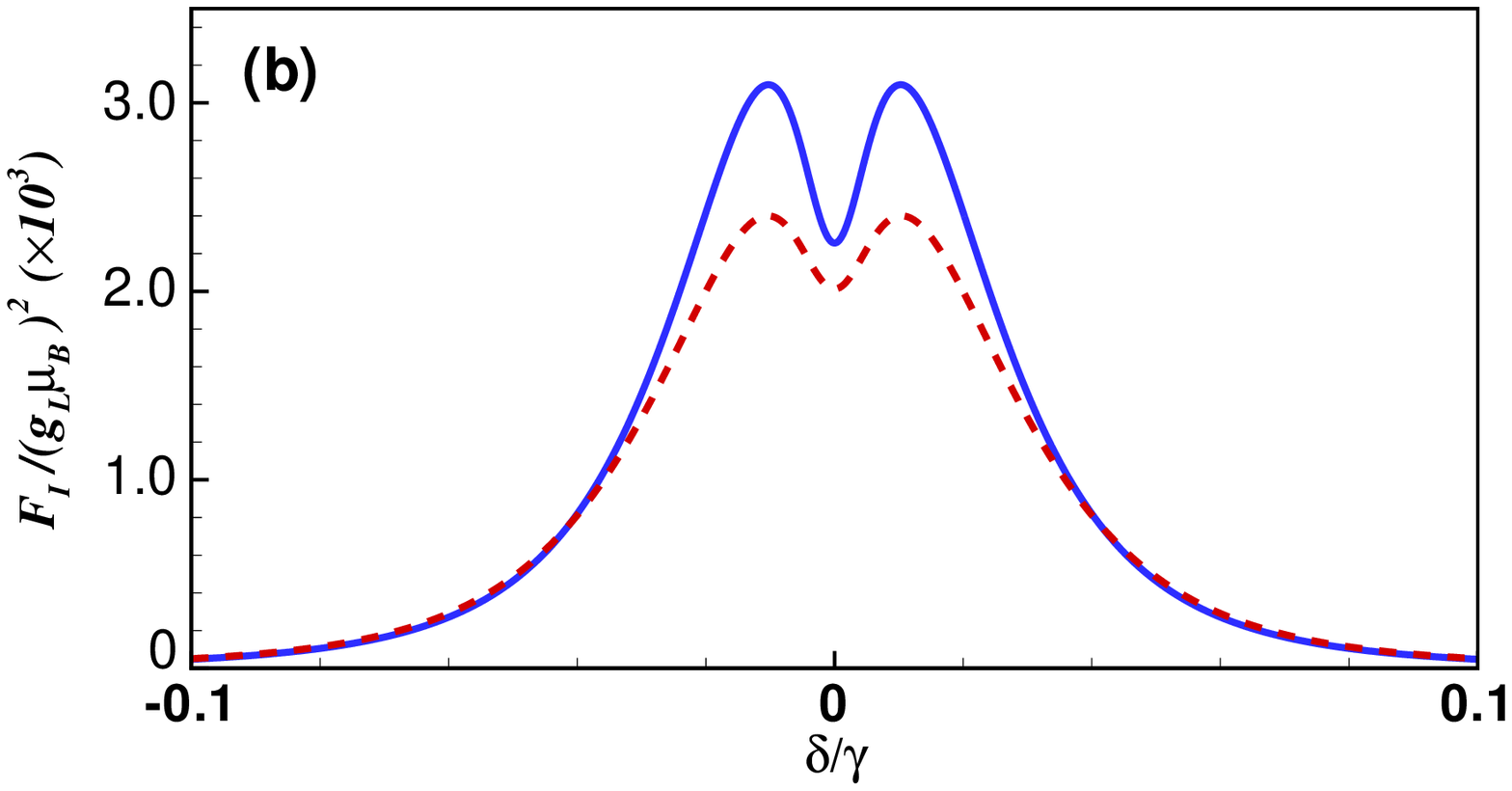}
\caption{(Color online) (a) The output probabilities $p(H|\delta)$
(dashed curve), $p(V|\delta)$ (solid curve), and $p(0|\delta)$ (long
dashed curve) versus the magnetic field induced level shift
$\delta$. (b) Fisher information as a function of $\delta$ without
(solid curve) and with (dashed curve) Doppler broadening. We take the Doppler width as $\Delta\omega_{D}\simeq0.56$~MHz (corresponding to $T=1$~mK). The relevant
parameters are taken as $\Delta_{p}=\Delta_{d}=0$, and other
parameters are same as those in Fig.~\ref{fig:tran-phase}.}\label{fig:fi}
\end{figure}

Assuming the input linear probe field is vertical polarized, the Faraday
rotation at the output can be measured by detecting the intensity
difference of two linearly polarized components. For each probe, the
output has three results: vertical polarized, horizontal polarized,
and no photon. The probabilities of three outputs are, respectively,
$p(H|\delta)=|t_{+}e^{i\phi_{+}}-t_{-}e^{i\phi_{-}}|^{2}/4$
($H$-polarized),
$p(V|\delta)=|t_{+}e^{i\phi_{+}}+t_{-}e^{i\phi_{-}}|^{2}/4$
($V$-polarized), and $p(0|\delta)=1-p(H|\delta)-p(V|\delta)$ (no
photon). We show the three outputs as functions of magnetic field
induced level shift $\delta$ in Fig.~\ref{fig:fi}(a). The probe
detuning is taken as $\Delta_{p}=0$, and all other relevant
parameters are the same as those in Fig.~\ref{fig:tran-phase}. Without the
magnetic field, the $\sigma_{+}$ and $\sigma_{-}$ components of the
probe field experience the same phase shift, and the polarization
rotation angle $\phi=0$. Thus no H-polarized photon can be detected
[see the dashed curve in Fig.\ref{fig:fi}(a)]. With increasing
of the magnetic field, $p(H|\delta)$ increases, and reaches it
maximum value $p(H|\delta)_{\rm max}=0.27(2)$ at
$\delta=\pm0.042(5)\gamma$. Increasing $\delta$ much more, no
transmitted photons can be detected.

The parameter $\delta$ can be estimated from the outputs. The
maximum information about weak magnetic field that can be extracted
from the measurement is given by the Fisher
information~\cite{zwierz-prl-2010}, which determines the limit of
the sensitivity. With single photon measurement, we
have~\cite{braunstein-prl-1994}
\begin{eqnarray}
{\rm S}\geq\frac{1}{\sqrt{\varsigma
F(\delta)}},\label{eq:sensitivity-singlephoton}
\end{eqnarray}
where $\varsigma$ is the number of times the procedure is repeated,
and the Fisher information $F(\delta)$ is given by
\begin{eqnarray}
F(\delta)=(g_{L}\mu_{B})^{2}\sum\limits_{x=
H,V,0}\frac{1}{p(x|\delta)}\left[\frac{\partial
p(x|\delta)}{\partial\delta}\right]^{2}.\label{eq:fi}
\end{eqnarray}

In practice, the atoms may move with velocity $v$, giving the $1/e$
temperature related Doppler width of the atomic ensemble as
$\Delta\omega_{D}=\sqrt{2k_{B}T\omega^{2}/(mc^{2})}$~\cite{geabanacloche-pra-2008}.
With $T=1.0$ mK, we immediately have
$\Delta\omega_{D}\simeq0.56$~MHz. Considering the influence of
Doppler
broadening, the Fisher information can be rewritten as
\begin{eqnarray}
F(\delta)=\frac{1}{\sqrt{\pi\Delta\omega_{D}^{2}}}
\int_{-\infty}^{\infty}F(\delta,kv)e^{-(kv)^{2}/\Delta\omega_{D}^{2}}d(kv),
\end{eqnarray}
in which $k$ is wave vector of the probe field. The Fisher
informations without (solid curve) and with (dashed curve) Doppler boradening versus level shift $\delta$ are depicted in Fig.~\ref{fig:fi}(b). All parameters are the same as those in
Fig.~\ref{fig:fi}(a). Confining the atomic ensemble in optical cavity, the detuning between the cavity mode and the probe field accuses an extra phase. While owing to the symmetry of the atomic configuration under consideration, we ignore the
systematic errors associated with ac Stark shifts~\cite{fleischhauer-pra-2000}. In the regime considered previously ($g\sqrt{N}\gg\Omega$), Equation (\ref{eq:tran}) show that the influence of the cavity shift on Fisher information can be safely neglected. With the realistic
experimental parameters ($^{87}$Rb $D_{1}$ line),
$\hbar\omega_{p}=1.55(9)$eV, $\gamma=\pi\times6.06$~MHz,
$\gamma'=10^{-3}\gamma=\pi\times6.06$~kHz, $g^{2}N=100\gamma^{2}$,
$\Omega=1\gamma$, $\kappa=2\gamma$, the sensitivity can be calculated
to be $2.3(1)$nT/$\sqrt{\rm Hz}$.

\begin{figure}
\includegraphics[width=8cm]{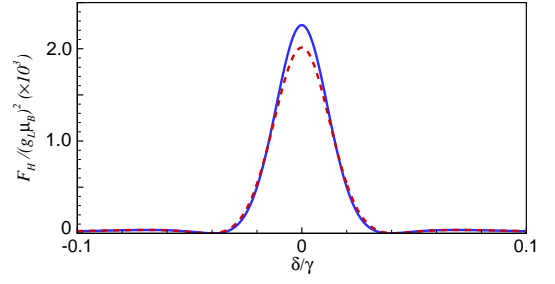}
\caption{(Color online) Fisher information of the horizontal polarized output as a function of $\delta$ without (solid curve) and with (dashed curve) Doppler broadening. All
parameters are same as those in Fig.~\ref{fig:fi}(b).}\label{fig:fi-h}
\end{figure}

Recalling the limit of small magnetic fields, and combining the
relation $g\sqrt{N}\gg\Omega$, the Fisher information $F(\delta)$
can be expanded as
\begin{eqnarray}
F(\delta)\simeq(g_{L}\mu_{B})^{2}\left[\frac{2}{\kappa^{2}}
\left(\frac{g\sqrt{N}}{\Omega}\right)^{4}+{\cal
O}(\delta^{2})\right].
\end{eqnarray}
The first dominating term demonstrates that the Fisher information
is proportional to $g^{2}N/\Omega^{2}$. Recalling the relation
(\ref{eq:sensitivity-singlephoton}), one immediately obtains ${\rm S}\sim1/N$, which
indicates that the sensitivity is inversely proportional to the
number of atoms confined in the cavity. This is the main result of
the collective coupling between the cavity modes and atomic
ensemble. It is therefore feasible to improve the sensitivity by
increasing the value of $g^{2}N/\Omega^{2}$. Taking
$g^{2}N=200\gamma^{2}$ and $\Omega=0.5\gamma$, the sensitivity can
be improved to $0.7(5)$nT/$\sqrt{\rm Hz}$ with single photon
measurement. It should be noted that a rise in atomic density is
accompanied by an increasing in dephasing rates between metastable
lower levels in such a system, and therefore optimal operating
conditions have to be carefully matched.

For multi photon measurement, following the
processes~\cite{xia-pra-2015}, the limit of the sensitivity can be
estimated by
\begin{eqnarray}
{\rm S}\geq \frac{t}{\sqrt{F_{\rm
H}}}\sqrt{\frac{\hbar\omega_{p}\sin^{2}\phi+2K_{B}T}{P_{\rm
in}}},\label{eq:sensitivity-multiphoton}
\end{eqnarray}
in which $F_{\rm
H}=(g_{L}\mu_{B})^{2}[\partial_{\delta}^{2}p(H|\delta)]^{2}/p(H|\delta)$
is the normal Fisher information of the horizontal polarized output,
and $K_{B}$ is the Boltzmann constant, $P_{\rm in}$ denotes the
input probe power. In the above derivation, the noise due to the internal loss channels are included.  Equation~(\ref{eq:sensitivity-multiphoton}) indicates that the
sensitivity is inversely proportional to $\sqrt{P_{\rm in}}$, and
thus can be improved by increasing the probe power. The normal
Fisher information of the horizontal polarized output is given by
\begin{eqnarray}
F_{H}=\frac{1}{\sqrt{\pi\Delta\omega_{D}^{2}}}\int_{-\infty}^{\infty}F_{H}(kv)e^{-(kv)^{2}/\Delta\omega_{D}^{2}}d(kv).
\end{eqnarray}
The evolution of $F_{H}$ versus $\delta$ with (dashed curve) and without (solid curve) Doppler broadening are shown in Fig.~\ref{fig:fi-h}. All parameters are the same as those in Fig.~\ref{fig:fi}(b). One obtains $F_{\rm H}\simeq2000(g_{L}\mu_{B})^{2}$~T$^{-2}$ (with Doppler broadening).
With $P_{\rm in}=1$mW, $T=1$~mK, the sensitivity can be
immediately calculated to be 16.7(8)~fT/$\sqrt{\rm Hz}$ with
$g^{2}N=100\gamma^{2}$ and $\Omega=1.0\gamma$. Changing the set of
parameters as $g^{2}N=200\gamma^{2}$ and $\Omega=0.5\gamma$, numerical calculations show that the sensitivity with 4.7(9)~fT/$\sqrt{\rm Hz}$ can be achieved with high
transmission [$t_{+}^{2}=t_{-}^{2}\simeq0.76(3)$].

\section{conclusions}

We have analyzed the cavity-enhanced optical Faraday rotation for weak
magnetic field detection in a cavity QED system consisting of
four-level atomic ensemble in tripod configuration confined in
cavity based on intracavity EIT. Owing to the collective coupling
between the cavity modes and atomic ensemble, the Faraday rotation
can be enhanced dramatically, and therefore leading to
cavity-enhanced sensitivity. With single photon measurement, the
sensitivity is inversely proportional to the number of the atoms,
and our numerical calculations show that the sensitivity with
0.7(5)~nT/$\sqrt{\rm Hz}$ can be achieved. While the sensitivity can
be improved by increasing the probe power with multiphoton
measurement, and sensitivity with 4.7(9)~fT/$\sqrt{\rm Hz}$ could be
attained. The cavity-enhanced Faraday rotation can be of interest in
atomic filter and quantum information processing.

\section*{ACKNOWLEDGMENT}

We would like to acknowledge useful discussion with Gongwei Lin.
This work was supported by the NSF Shaanxi Province (2014JM2-1001), as well as the Central
University (GK201505034).

\end{document}